\documentclass{ortex}
\usepackage{graphicx}
\usepackage{url}
\title{Design study of the KAGRA output mode-cleaner}

\author{Ayaka Kumeta\thanks{E-mail address: kumeta@gw.phys.titech.ac.jp},
Charlotte Bond$^{1}$,
and Kentaro Somiya
}
\inst{Department of Physics, Tokyo Institute of Technology, 2-12-1 Oh-okayama, Meguro-ku, Tokyo 152-8551, Japan\\
$^{1}$School of Physics and Astronomy, University of Birmingham, Edgbaston, Birmingham B15 2TT, United Kingdom
}
\abst{Most second-generation gravitational-wave detectors employ an optical resonator called an {\it output mode-cleaner (OMC)}, which filters out junk light from the signal and the reference light, before it reaches the detection photodiode located at the asymmetric port of the large-scale interferometer. The optical parameters of the OMC should be carefully chosen to satisfy the requirements to filter out unwanted light whilst transmitting the gravitational wave signal and reference light. The Japanese gravitational-wave detector KAGRA plans to use a very small amount of reference light to minimize the influence of quantum noise for gravitational waves from binary neutron stars, and hence the requirements to the OMC are more challenging than for other advanced detectors. In this paper, we present the result of numerical simulations, which verify that the OMC requirements are satisfied with the current design. We use the simulation program {\it FINESSE} and realistic mirror phase maps that have the same surface quality as the KAGRA test masses.}

\kword{Interferometer, Simulation, Gravitational waves}

\begin{document}
\maketitle

\section{Introduction}\label{Sec:1}

\begin{figure}
\begin{center}
\includegraphics[width=7.5cm]{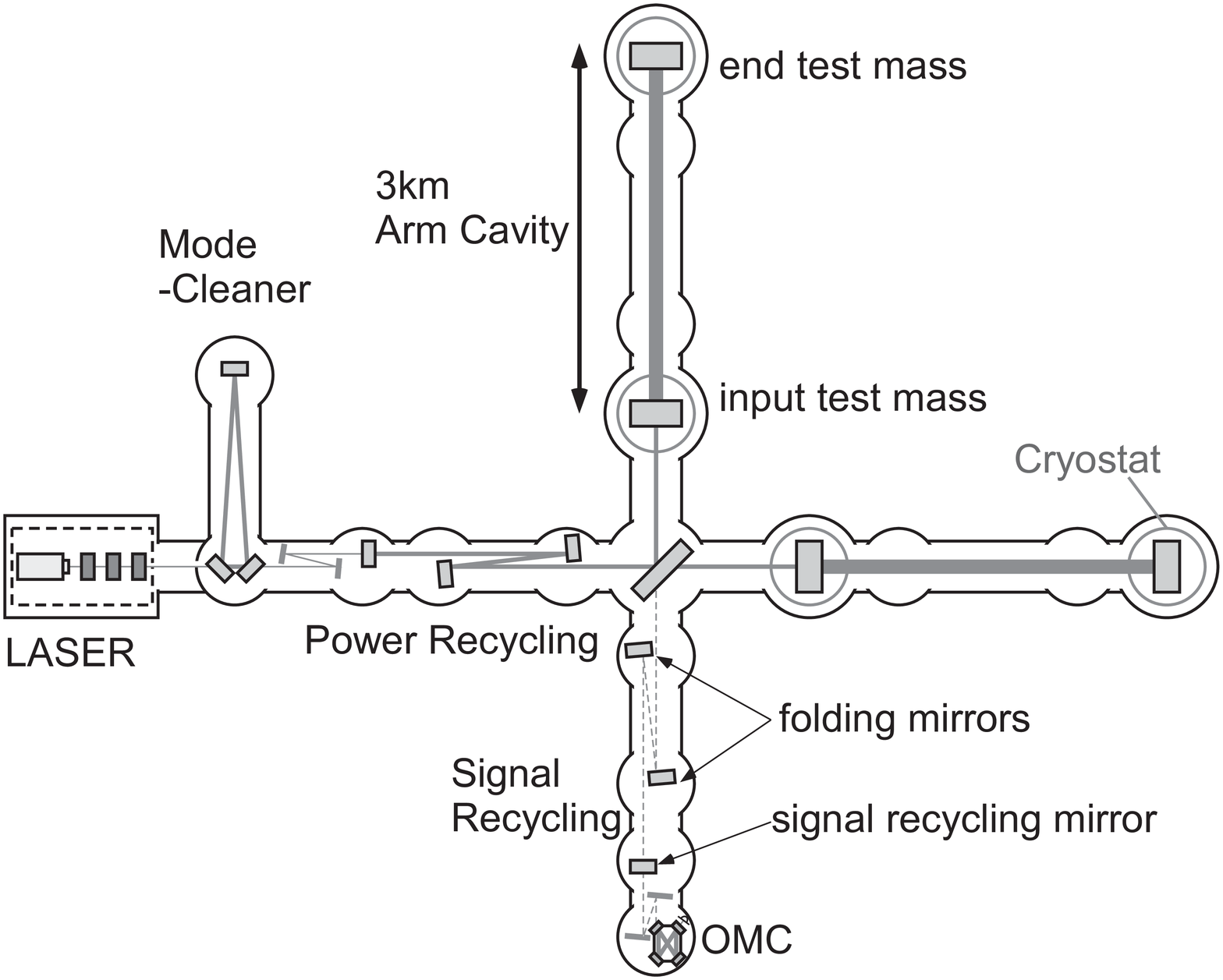}
\hspace{0.5cm}
\includegraphics[width=6cm]{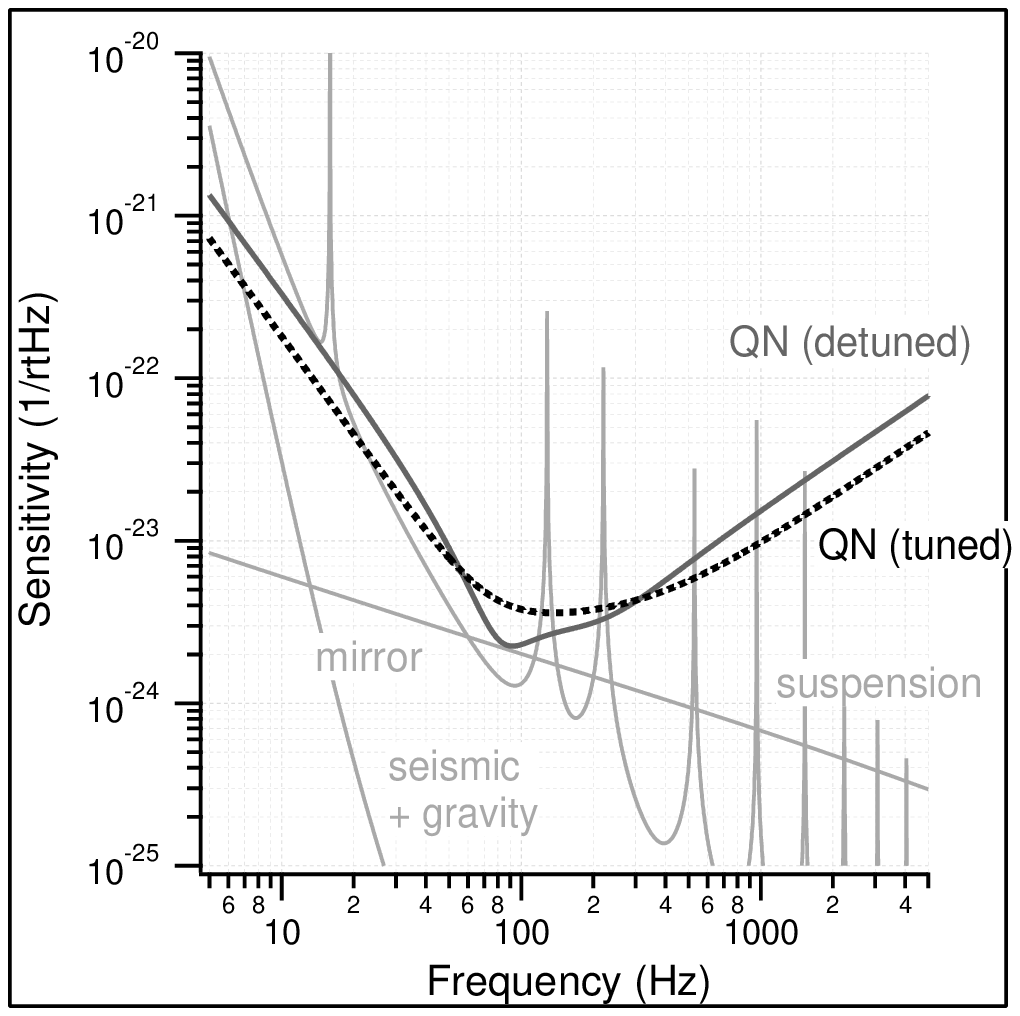}
\end{center}
\caption{{\it Left}: Interferometer configuration of KAGRA, which contains 11 mirrors in the main interferometer including the input/end test masses and power/signal recycling mirrors. The OMC is located at the anti-symmetric port after the signal-recycling mirror where the gravitational-wave signal comes out. {\it Right}: Noise budget of KAGRA. The quantum noise spectrum changes with the detuning of the signal recycling cavity and the homodyne readout phase. The total sensitivity is given by the square sum of the noise curves.}
\label{fig:IFO}
\end{figure}

Gravitational waves are ripples of space-time predicted by Einstein~\cite{Einstein}. The historical first detection of a gravitational wave is expected to be realised with the second-generation detectors that will begin taking science data over the next few years. KAGRA is the Japanese second-generation detector being constructed underground in the Kamioka mine~\cite{SomiyaCQG}. The detector is based on a Michelson interferometer locked on the dark fringe, with Fabry-Perot optical resonators in each arm and additional resonators in the symmetric and anti-symmetric ports of the interferometer (see Fig.~\ref{fig:IFO}, left panel). The layout and parameters of the interferometer are carefully chosen to enhance the effect of a gravitational wave and suppress sources of noise in the detector. The mirrors in the arm cavities are made of sapphire and cooled down to 20\,K to reduce thermal noise. The underground facility and cryogenic operation are the two main characteristics of KAGRA. With the subsequent reductions in seismic noise and thermal noise, quantum noise limits the sensitivity in the observation band (see Fig.~\ref{fig:IFO}, right panel). The primary target of KAGRA is gravitational waves from binary neutron stars~\cite{NS}. For such signals, the gravitational-wave frequency will gradually increase until $\sim$1.5\,kHz when the two stars collide. Since the noise level is high at low frequencies and the signal is weak at high frequencies, the detector is most sensitive to the inspiral signal at around 100\,Hz and it is the most efficient to improve the sensitivity at around the frequency. The observation range will be around $120\sim150$\,Mpc depending on the operation of the signal recycling cavity, which is explained in Sec.~\ref{Sec:2}.

\section{Signal readout scheme}\label{Sec:2}

While the optical resonator at the symmetric port of the interferometer, called the {\it power-recycling cavity}, resonates the carrier light to increase the input power to the interferometer, the role of the optical resonator at the anti-symmetric port, called the {\it signal-recycling cavity}, is to resonate the gravitational waves signal sidebands exiting the Michelson interferometer to optimize the sensitivity spectrum~\cite{Mizuno}. The signal-recycling cavity length can be detuned from the resonance or anti-resonance of the carrier light in order to maximize gravitational wave signals from binary neutron stars expected at specific frequencies, and so maximize the observation range for such events. However, this can significantly decrease the sensitivity for other sources if the detuning is too large. 

\begin{figure}
\begin{center}
\includegraphics[width=0.6\linewidth]{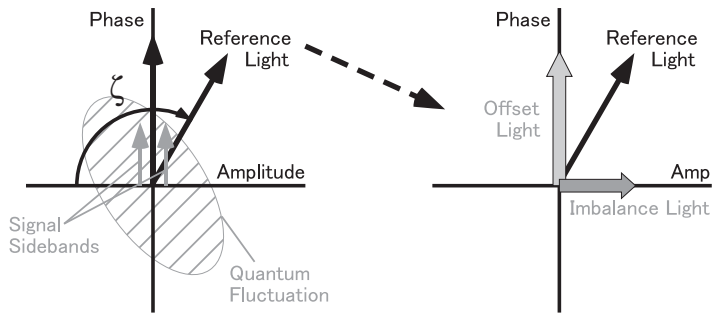}
\end{center}
\caption{{\it Left}: Phasor diagram of the fields at the signal extraction port in the case without detuning. Gravitational waves create the upper and lower signal sidebands. The upper rotates counter-clockwise and the lower rotates clockwise in the diagram. The sum of the two sidebands is always in the phase quadrature. Quantum fluctuations consists of shot-noise and radiation-pressure-noise components. The readout phase $\zeta$ should be carefully chosen to optimize the signal-to-noise ratio at the most important signal frequencies. {\it Right}: The reference light consists of light leaking into the dark port from reflectivity imbalances between the two arm cavities and light leaking through from the differential offset of the operating point of the arm cavities. While the former is fixed due to mirror imperfections, the latter can be controlled to select the optimal readout phase.}
\label{fig:phasor}
\end{figure}

A gravitational wave will modulate the phase of the carrier light in the arm cavities to create the upper and lower sidebands at the gravitational-wave frequency. Without the detuning of the signal-recycling cavity, the signal sidebands are balanced; in the phasor diagram that expresses the relative phase difference from the carrier light, the sum of the sidebands is always in the phase quadrature (see Fig.~\ref{fig:phasor}, the left panel). While the signal readout scheme used in first-generation detectors was a radio frequency heterodyne readout scheme~\cite{RF}, second-generation detectors aim to use a homodyne readout scheme using carrier light leaked into the dark port as a reference light to beat with gravitational wave signals, which is called {\it DC readout}~\cite{DC}. There are two ways in which some of the carrier light in the arm cavities can leak into the dark port. One is via a reflectivity imbalance between the two arm cavities. This component appears in the amplitude quadrature in the phasor diagram. The other is via differential offsets to the operating point of the arm cavities. This component appears in the phase quadrature as a constant signal at DC. The combination of the two components determines the readout phase of the reference field (see Fig.~\ref{fig:phasor}, the right panel). In the tuned configuration, the gravitational-wave signal is extracted in the most efficient way with a DC readout phase of 90\,deg. However, quantum radiation pressure noise is high in the phase quadrature and so the readout phase should be optimized for the target source even for zero detuning of the signal recycling cavity. This becomes more complicated in a detuned configuration. The upper and lower signal sidebands are no longer balanced. To select the readout phase, one should compare the spectra and observation ranges for different combinations of detuning and readout phase. For KAGRA, a detune phase of 3.5\,deg has been selected, with a corresponding DC readout phase of 132\,deg~\cite{SomiyaCQG}. The observation range for zero detuning and with a readout phase of 90\,deg is 128\,Mpc while that with 3.5\,deg detuning and the optimal readout phase is 148\,Mpc. The optimization of the readout phase is important in KAGRA where the sensitivity is limited by quantum noise.

\section{Output mode-cleaner}\label{Sec:3}

\subsection{Requirements}
The laser beam propagating in the interferometer can be described as a sum of the Hermite-Gauss beams. In this study, the fundamental mode shall be set as the nearest fundamental mode of the signal field and higher order Gauss beams are regarded as junk light. The OMC plays an important role in filtering out the higher order spatial modes that are generated in the arm cavities and the radio frequency (RF) sidebands used for the interferometer control, before the output light reaches the detection photodiode. In this study, 515\,W is assumed for the laser power at the beam splitter~\cite{SomiyaCQG}. The radius of curvature of the test masses is 1.9\,km and we assume $\pm1\,\%$ error between the two arm cavities. The arm cavity finesse is 1550 and the roundtrip optical loss is 100\,ppm. Here we also assume a $\pm10\,\%$ loss imbalance between the two arm cavities. The roundtrip Gouy phase shifts in the power and signal recycling cavities are 33\,deg and 35\,deg, respectively~\cite{Aso}, and the incident angles at the folding mirrors in the recycling cavities are about 0.6\,deg. The power reflectivity of the signal-recycling mirror is 85\,\%.

\begin{figure}[ht]
\begin{center}
\includegraphics[width=0.5\linewidth]{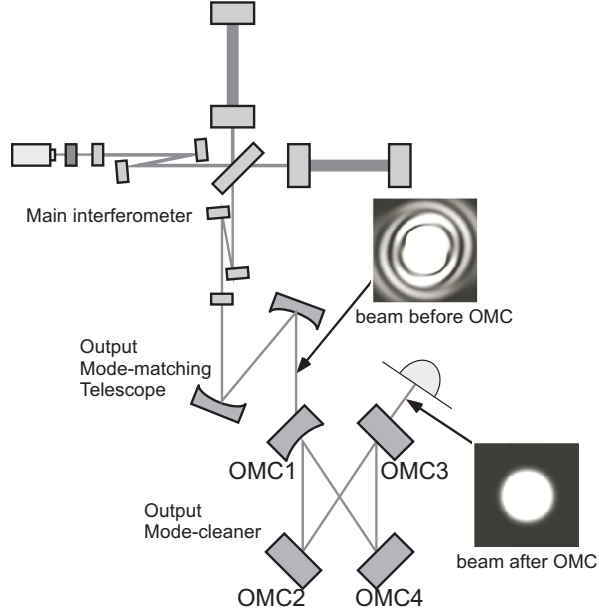}
\end{center}
\caption{The OMC filters out junk light and lets a gravitational wave signal and the reference light for the DC readout transmit through the detection photodiode. The OMC consists of 4 mirrors: the first mirror (OMC1) is curved and the others are flat.}
\label{fig:OMC}
\end{figure}

As a result of a simulation without a mirror phase map included, the power of the DC light that is to be used as reference light is about 1\,mW before the OMC, achieved by applying a differential offset to the arms of the main interferometer.
The KAGRA OMC must not degrade the shot-noise level by more than 5\,\% compared with the case of an ideal OMC that filters out all the unnecessary light. If the DC light power is 1\,mW, a possible breakdown of an acceptable degradation after the OMC is as follows; the signal loss is to be 2\,\% or less, the residual spatial higher order modes are to be 10\,$\mu$W or less, and the residual RF sidebands are to be 20\,$\mu$W or less.
This requirement can be satisfied by selecting a proper parameter set; namely, the finesse, round-trip Gouy phase, and length of the OMC.

\subsection{Finesse, Gouy phase, and cavity length}
The finesse of an optical cavity is a number which represents how many times the light circulates in the cavity and is determined by the reflectivity of the cavity mirrors. In the case of a 4-mirror ring cavity with mirror reflectivities $r_1$, $r_2$, $r_3$, and $r_4$, the finesse is given by
\begin{eqnarray}
 {\cal F} = \frac{\pi \sqrt{r_1r_2r_3r_4}}{1-r_1r_2r_3r_4}\ .
\end{eqnarray}
To maximize transmission of the signal through the OMC we use equally partially transmissive mirrors OMC1 and OMC3 (i.e. $r_1=r_3<1$) and nearly perfectly reflective mirrors for OMC2 and OMC4 ($r_2=r_4\simeq1$).
The optical loss ${\cal L}$ for each OMC mirror is expected to be about 40\,ppm per bounce~\cite{Arai2}. If the finesse is too high, the light circulates too many times and the total optical loss of the signal would not satisfy the requirement. In order to keep the signal loss at less than 2\,\%, the finesse should not be higher than 800.

\begin{figure}
\begin{center}
\includegraphics[width=8cm, bb= 150 400 600 700]{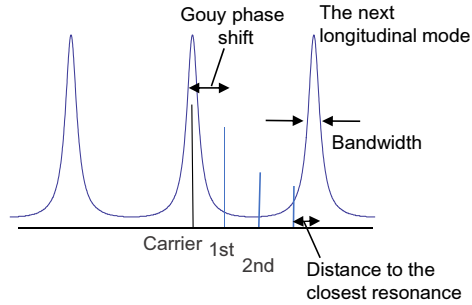}
\end{center}
\caption{Resonant condition of different light fields in a cavity. The carrier is on resonance and the higher order spatial modes are off resonance due to the Gouy phase shift. The sharpness of the resonance peak increases with finesse and the spacing between subsequence peaks decreases with cavity length.}
\label{fig:proper}
\end{figure}

The OMC filters out higher order spatial modes, with the filtering efficiency of each mode determined by the round-trip Gouy phase of the OMC cavity or by the radius of curvature of the OMC mirrors:
\begin{eqnarray}
 {\eta = \arccos \sqrt[]g}\hspace{1cm}
\left(g = 1 - \frac{L}{R}\right)
\end{eqnarray}
Here $g$ is the g-factor of the cavity, $L$ is half the roundtrip length of the OMC cavity and $R$ is the radius of curvature of the curved mirror (note that only OMC1 is curved whilst the other mirrors are flat).
The OMC power transmittance, of the m-th spatial mode is given by 
\begin{eqnarray}
 T_\mathrm{omc} =\frac{(1-r_1^2-{\cal L})(1-r_3^2-{\cal L})r_2^2r_4^2}{\left(1-r_1r_2r_3r_4\mathrm{e}^{-i\mathrm{m}\eta}\right)^2}\ .
\end{eqnarray}
%Here $r$ is the reflectivity of each mirror, $\cal L$ is the optical loss of each mirror and $j$ is the number of each mirror.
If the Gouy phase shift accumulated by a higher order mode is close to an integer multiple of 2$\pi$, the m-th mode nearly resonates in the OMC and cannot be filtered out. The ratio of the distance to the closest resonance to the cavity band-width determines the filtering efficiency for each higher order mode (Fig.~\ref{fig:proper}). It is important to select a proper combination of the Gouy phase and the cavity length, paying attention to the expected power in each higher order mode before the OMC, which should be numerically calculated using a simulation tool (see Sec.~\ref{Sec:4}).

The power of an RF sideband transmitted by the OMC is given by 
\begin{eqnarray}
 T_\mathrm{RF} =\frac{(1-r_1^2-{\cal L})(1-r_3^2-{\cal L})r_2^2r_4^2}{\left(1-r_1r_2r_3r_4\mathrm{e}^{-i\mathrm{m}\eta+2L\omega_\mathrm{RF}/c}\right)^2}\ .
\end{eqnarray}
Here $\omega_\mathrm{RF}$ is the radio frequency. In the case of KAGRA, $\pm16.875$\,MHz sidebands exit the anti-symmetric port of the interferometer.
In order to filter out the RF sideband to the requirement of less than 20\,$\mu$W after the OMC, $L$ should be longer than 75\,cm.

\section{Simulation with mirror maps}\label{Sec:4}

\subsection{FINESSE}
\begin{figure}[htb]
\begin{center}
\includegraphics[width=7cm]{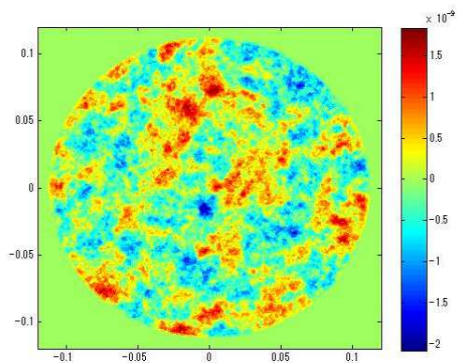}\\
\end{center}
\caption{One of the 4 mirror maps used in the simulations detailed here. For all the mirror maps, the mirror diameter is 220\,mm and the rms (root-mean-square) surface error is 0.5\,nm.}
\label{fig:map}
\end{figure}

FINESSE is a simulation tool for modelling interferometers based on the modal model{\cite{finesse}\cite{finesse2}}. FINESSE can analyze waves propagating in a user-defined interferometer not only for a plane wave approximation but also using a Hermite-Gauss modal expansion to describe the shape and spatial properties of light fields in a given setup. In this paper, we use FINESSE to simulate the expected beam distortions due to the nanometer-scale mirror surface errors of the test masses. Figure~\ref{fig:map} shows one of the mirror maps used in the simulations documented here.

\subsection{Mirror maps}
A mirror phase map is an array of numbers representing the height of a mirror over its surface. The phase maps were generated so that their PSDs (power spectral density) are consistent with the KAGRA mirror PSD specification, determined based on the loss requirement~\cite{yamamoto_san}. We create 4 independent mirror maps with the same PSD and apply them to the 4 test masses of the interferometer. There are 24 combinations and the results provide a statistical distribution of the higher order modes exiting the interferometer. 

\subsection{Selection of the optical parameters for the KAGRA OMC}
\begin{figure}[htb]
\begin{center}
\includegraphics[width=0.48\linewidth]{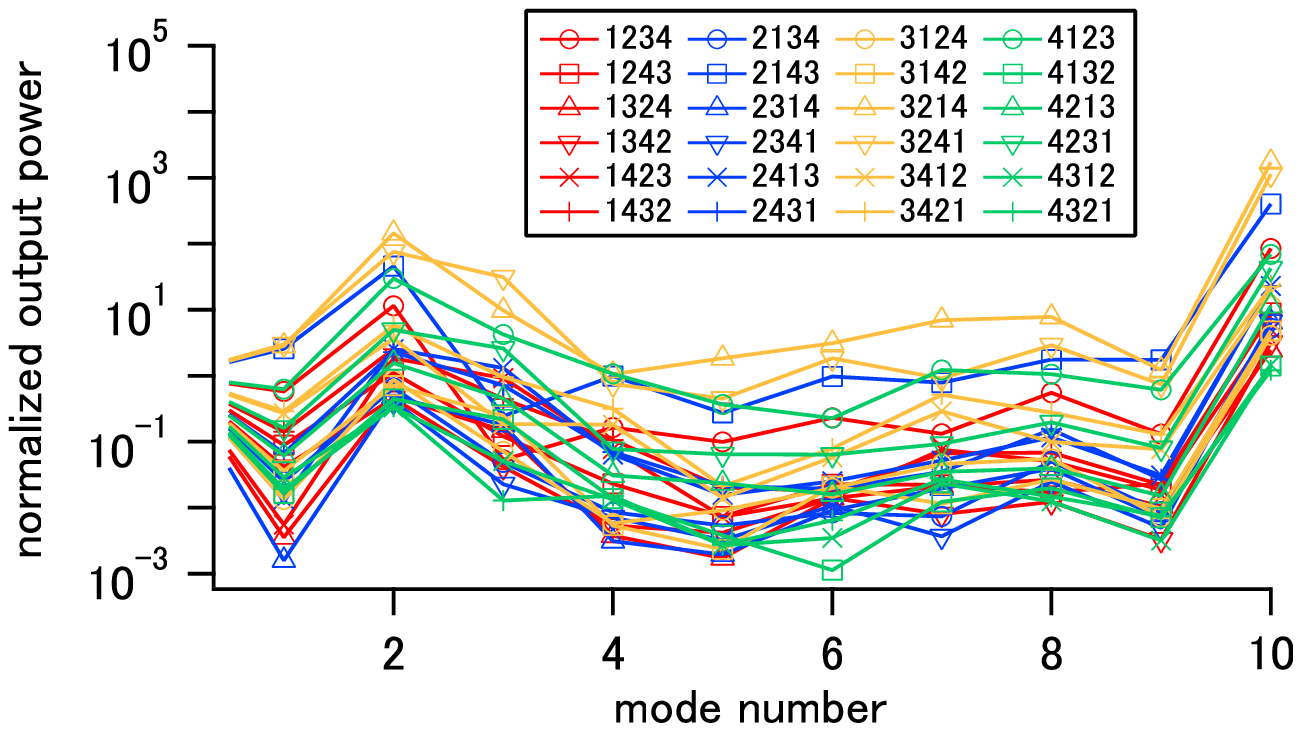}
%\hspace{0.2cm}
\includegraphics[width=0.48\linewidth]{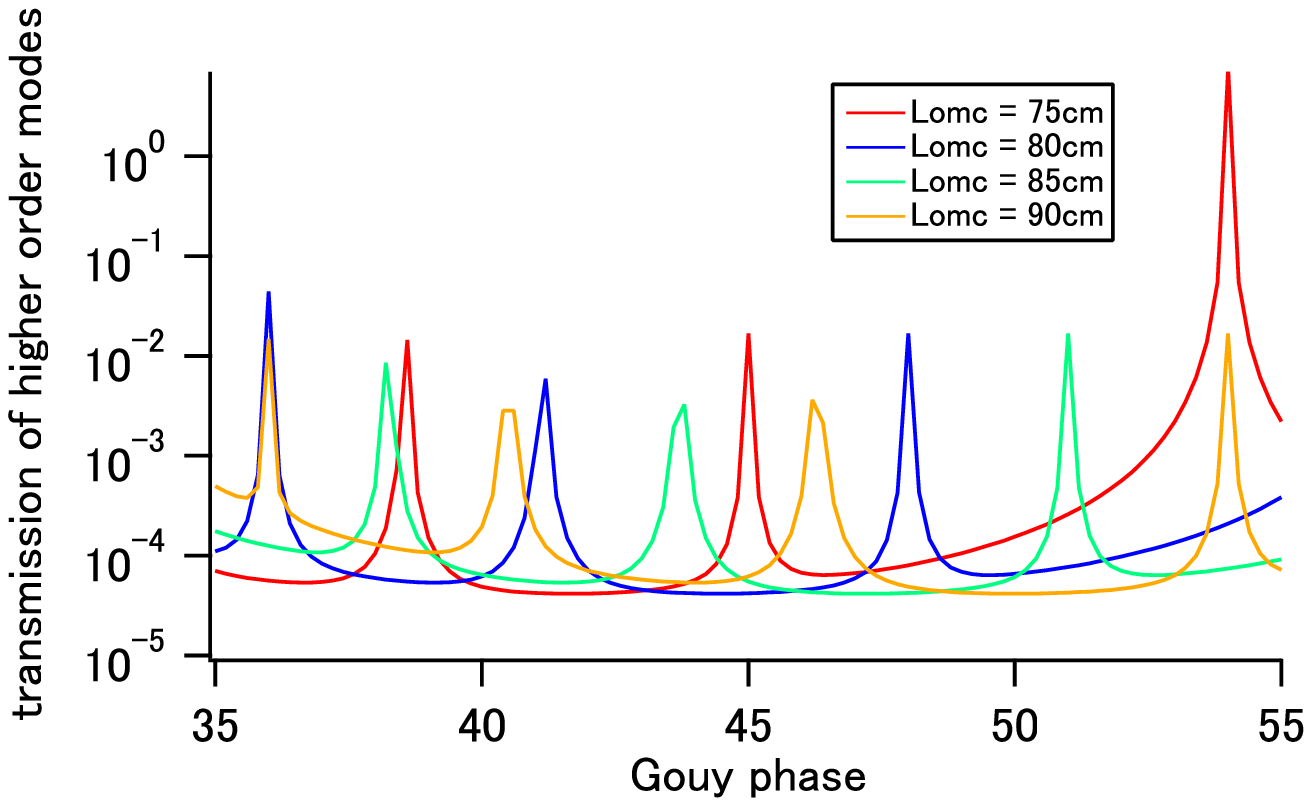}
\end{center}
\caption{{\it Left}: The normalized output power of the m-th mode beam $(\mathrm{m}\geq1)$ for each of the 24 combinations of mirror maps. {\it Right}: Transmitted power of each higher order mode multiplied by a weighting factor calculated from the simulations of the interferometer output.}
\label{fig:10ji}
\end{figure}

\begin{figure}[htb]
\hspace{1cm}
 \begin{minipage}{0.5\hsize}
 \begin{center}
 \includegraphics[width=5cm]{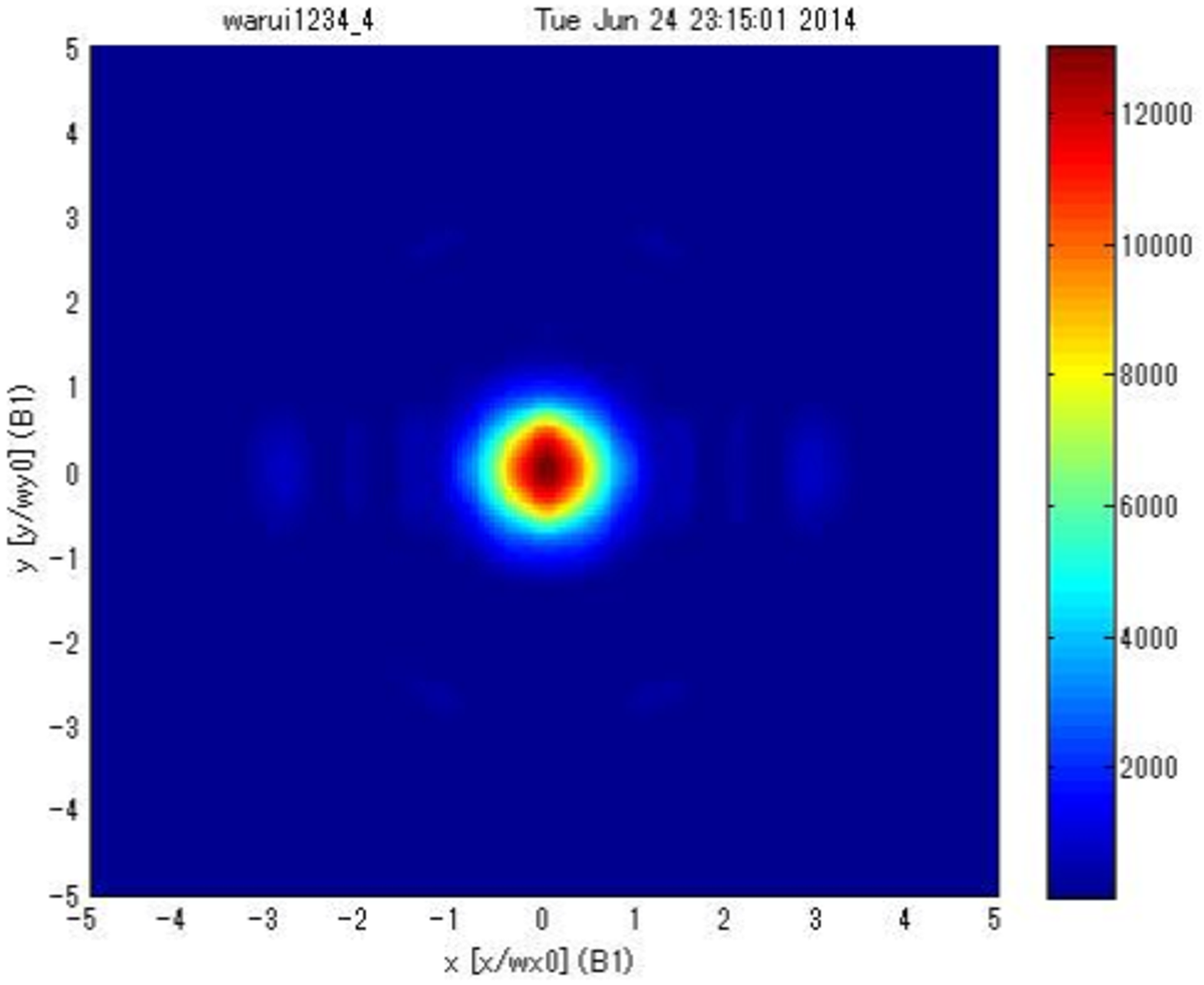}\\
 \end{center}
 \end{minipage}
\hspace{-1.5cm}
 \begin{minipage}{0.5\hsize}
 \begin{center}
 \includegraphics[width=5cm]{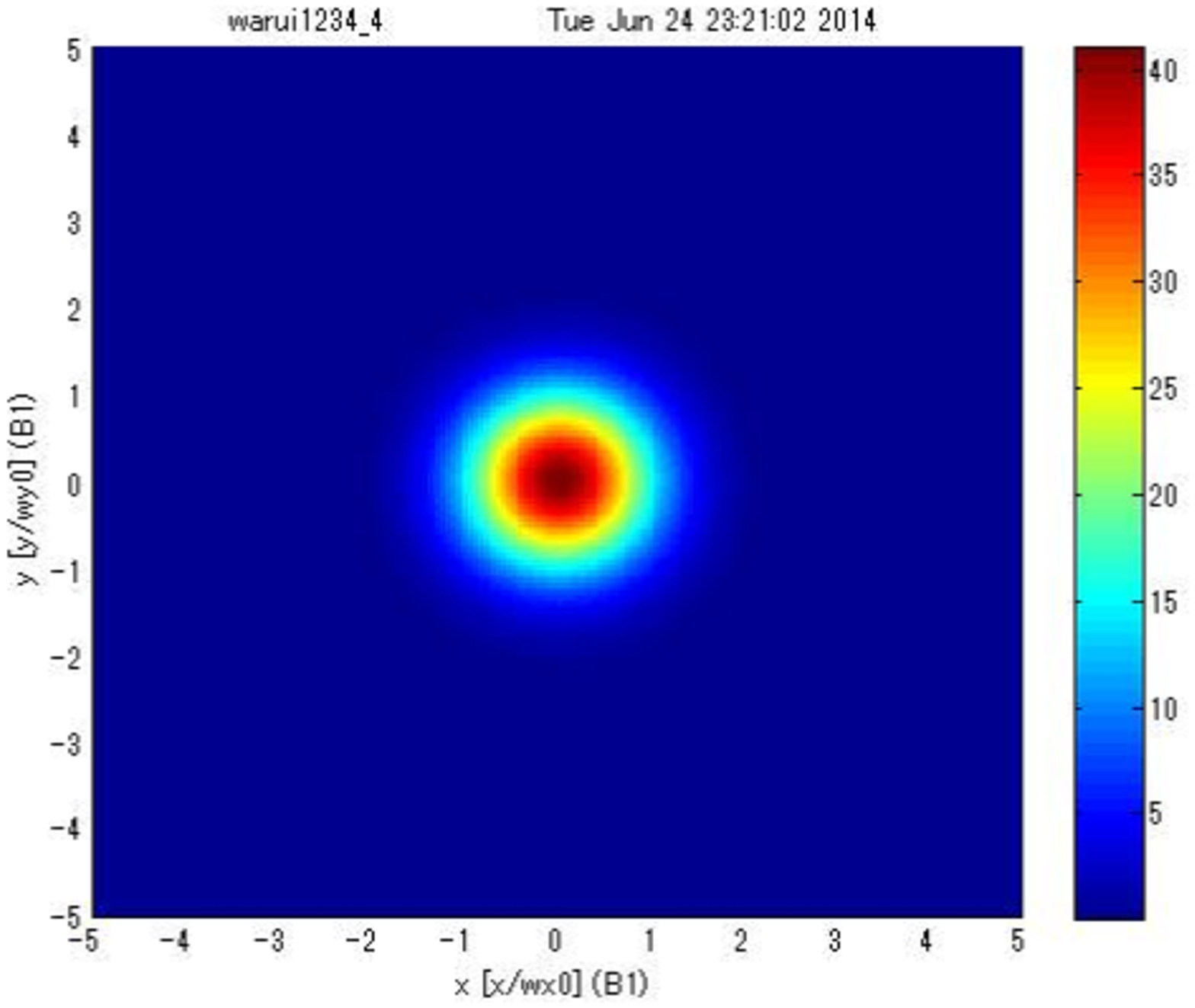}\\
 \end{center}
 \end{minipage}
 \caption{Profile of the beam before (left) and after (right) the OMC. Here the RF sidebands are excluded.}
\label{fig:before_after}
\end{figure}

In a past study, Waldman calculated the distance between each higher order mode and the nearest OMC resonance for the first 8 modes to select the most appropriate optical design of the OMC~\cite{Sam}. While Waldman treated all the higher order modes equally, Arai implemented a weighting factor on each mode according to the beam distortions observed in a similar interferometer~\cite{Arai}. For the KAGRA OMC, we can predict proper weighting factors according to simulation results with realistic mirror maps. The left panel of Fig.~\ref{fig:10ji} shows the total power of the m-th mode at the output of the interferometer before the OMC, normalized by the power in the fundamental mode for the 24 different mirror map combinations. In KAGRA, the 10th mode comes out to the anti-symmetric port the most since its distance to the neighboring resonance is as small as 6\,deg, which is the closest among the first 20 modes. We take the average of these values and use these as the weighting factors. We then multiply the weighting factors by the OMC transmittance $T_\mathrm{omc}$ for each mode. The results are compared with different Gouy phases and cavity lengths and we find that $\eta=45$\,deg and $L=80$\,cm is the most appropriate parameter set for the KAGRA OMC (Fig.~\ref{fig:10ji}, the right panel). As is shown in Table~\ref{Table:BeforeAfter} and Fig.~\ref{fig:before_after}, the higher order modes and the RF sidebands are well suppressed with this KAGRA OMC design.

\begin{table}[htb]
\begin{tabular}{|c|c|c|c|c|c|c|} \hline
   & 0th & 1st & 2nd & 3rd & 10th & RF \\ \hline
   Before OMC & 877\,$\mu$W & 504\,$\mu$W & 10.1\,mW & 46.5\,$\mu$W & 74.2\,mW & 409\,mW \\ \hline
   After OMC & 853\,$\mu$W & 4.34\,nW & 39.8\,nW & 298\,pW & 340\,nW & 20.6\,$\mu$W \\ \hline\end{tabular}
\caption{The power in higher order modes and RF sidebands, before and after the OMC.}
\label{Table:BeforeAfter}
\end{table}

In the end, we calculate the shot noise limited sensitivity of KAGRA and compare the results for a lossless high-finesse OMC and with the realistic KAGRA OMC. For the comparison, we use the 5th worst combination of the mirror maps since the fundamental mode beam power is less than 0.2\,mW in the cases with even worse combinations (in such unlucky cases, we will have to increase the DC offset, descoping the readout strategy). The difference of the shot noise limited sensitivities is 3.2\,\% at 100\,Hz, which satisfies the requirement of the KAGRA OMC.

\section{Summary}

We performed a modal model simulation to design the output mode-cleaner (OMC) for KAGRA. The requirement for the OMC is not to degrade the sensitivity by more than 5\,\% due to higher order spatial modes leaking through from the interferometer due to the surface errors of the mirrors, transmitted RF sideband used for auxiliary controls, and the optical loss of the OMC itself. For this study, we used 24 combinations of 4 independent mirror maps that are created with a statistic random function and a realistic spatial power spectrum. The simulation result provided the contribution of each higher order mode before the OMC and we used this information to choose a proper set of optical parameters for the OMC. We then picked the 5th worst combinations of mirror maps to calculate the shot noise limited sensitivity and found that the KAGRA OMC satisfies the requirement.

It turned out that the DC light power could be less than 0.2\,mW in 4 out of the 24 cases we tested. In such unlucky cases, it is not possible to satisfy the requirement: the sensitivity would be deteriorated either from the contribution of the RF sidebands or from the optical loss if we further increase the finesse. If the sensitivity degradation is more than 20\,\%, it may be necessary to change the readout strategy; increasing the DC offset, to try to maintain the sensitivity at specific frequencies (for binary inspirals), and possibly sacrifice the sensitivity at other frequencies.

\acknowledgment
The authors would like to thank Daniel Brown and Andreas Freise for their kind help on the use of FINESSE, and Hiro Yamamoto for his kind support in creating realistic mirror maps. Special thanks also go to Jumpei Kato, Kazushiro Yano, Sho Atsuta, Yu Kataoka for their kind support. This work is partially supported by the {\it Specially Promoted Research} and the {\it Core-to-Core Program} of the Japan Society for the Promotion of Science (JSPS).

%\appendix
%\section{}

\end{document}